\documentclass[twocolumn,showpacs,preprintnumbers,amsmath,amssymb,superscriptaddress,pra]{revtex4}
\usepackage[latin1]{inputenc}
\usepackage{dcolumn}
\usepackage{bm}
\usepackage{graphicx}

\def\bbm[#1]{\mbox{\boldmath $#1$}}
\newcommand{\ket}[1]{\displaystyle{|#1\rangle}}
\newcommand{\bra}[1]{\displaystyle{\langle #1|}}
\newcommand{\sumpk}{\sum_p\int_0^{+\infty}dk_z\int d^2\mathbf{k}}
\newcommand{\sumpb}{\sum_{p=1}^{+\infty}\sum_{b=1}^{+\infty}}
\newcommand{\apk}{a_p(\mathbf{k},k_z)}
\newcommand{\acpk}{a^\dag_p(\mathbf{k},k_z)}
\newcommand{\Apkz}{A^{(1)}_p(\mathbf{k},k_z,\mathbf{r})}

\newcommand{\fpkz}{\mathbf{f}_p(\mathbf{k},k_z,z)}

\newcommand{\rv}{\mathbf{r}}
\def\TE{\mathrm{TE}}
\def\TM{\mathrm{TM}}

\begin{document}

\title{Atomic states in optical traps near a planar surface}
\author{Riccardo Messina}\email{riccardo.messina@obspm.fr}
\author{Sophie Pelisson}\author{Marie-Christine Angonin}
\author{Peter Wolf}\affiliation{LNE-SYRTE, Observatoire de Paris, CNRS UMR8630,
UPMC\\61 avenue de l'Observatoire, 75014 Paris, France}

\date{\today}

\begin{abstract}
In this work we discuss the atomic states in a vertical optical
lattice in proximity of a surface. We study the modifications to
the ordinary Wannier-Stark states in presence of a surface and we
characterize the energy shifts produced by the Casimir-Polder
interaction between atom and mirror. In this context, we introduce
an effective model describing the finite size of the atom in order
to regularize the energy corrections. In addition, the
modifications to the energy levels due to a hypothetical
non-Newtonian gravitational potential as well as their
experimental observability are investigated.
\end{abstract}


\maketitle

\section{Introduction}

Atomic interferometry has the potential to become a powerful
method to investigate atom-surface interactions, the main reason
being the high precision which can be reached in frequency
measurements. In this context a new experiment named FORCA-G
(FORce de CAsimir et Gravitation à courte distance) has been
recently proposed \cite{WolfPRA07}. The purpose of this experiment
is manifold: on one hand it aims at providing a new observation of
the Casimir-Polder interaction between an atom and a surface,
resulting from the coupling of the fluctuating quantum
electromagnetic field with the atom \cite{ScheelActaPhysSlov08};
on the other hand it also intends to impose new constraints on the
existence of hypothetical deviations from the Newtonian law of
gravitation. These goals will be achieved thanks to the innovative
design of FORCA-G, in which interferometric techniques are
combined with a trapping potential. This is generated by a
vertical standing optical wave produced by the reflection of a
laser on a mirror. The vertical configuration leads to an external
potential on the atom given by the sum of the optical one and a
linear gravitational term due to the earth: this deviation from a
purely periodical potential produces a localization of the atomic
wavepacket, corresponding to the transition from Bloch to
Wannier-Stark states \cite{GluckPhysRep02}. The main advantages of
FORCA-G are thus the refined control of the atomic position as
well as the high precision of interferometric measurements, as
demonstrated in the first experimental results
\cite{BeaufilsArxiv}.

Having in mind a theory-experiment comparison within a given
accuracy, the theoretical treatment of the problem as well as the
experimental investigation must be independently assessed with the
same precision. In the case of FORCA-G this demands a detailed
theoretical study of the atomic wavefunctions and energy levels in
proximity of a surface. As an intermediate step, a precise
characterization of the Casimir-Polder atom-surface interaction is
also needed.

These issues are the main subject of investigation of this paper.
As a matter of fact, the influence of the Casimir-Polder
interaction on the atomic energy levels has so far been explored
\cite{WolfPRA07,DereviankoPRL09} using the simple idea of
calculating the electrodynamical potential at the center of each
well of the trap. We will discuss the validity of this model
focusing in particular on the scheme of FORCA-G. In this work, we
present a hamiltonian approach to this problem. This treatment
first allows us to discuss, independently on the Casimir-Polder
atom-surface interaction, the atomic trapped states. Since the
presence of the surface breaks the translational symmetry typical
of Bloch and Wannier-Stark problems, we focus in particular on the
difference (both in energy levels and wavefunctions) between our
trapped states and the standard Wannier-Stark solutions. Then, in
order to discuss the Casimir-Polder corrections to the energy
levels, we generalize the perturbative treatment usually exploited
to deduce atom-surface electrodynamical interactions, by including
the external optical and gravitational potentials, and treating as
a consequence the atomic coordinate as a dynamic variable. The
theoretical work presented here will be useful for all experiments
that aim at measuring short range interactions between atoms
trapped in optical lattices and a macroscopic surface
\cite{WolfPRA07,DereviankoPRL09,SorrentinoPRA09} as they will
require precise modelling of the atomic states and energy levels
close to the surface.

This paper is organized as follows. In section \ref{Sec:2} we
describe our physical system. Then, in section \ref{Sec:3} we
discuss the shape of atomic wavefunctions in the trap. Section
\ref{Sec:4} is dedicated to the study of the Casimir-Polder
interaction and its influence on the atomic energy levels. In this
section we introduce an effective description of the finite size
of the atom and discuss its validity in connection with the
experiment. In section \ref{Sec:5} we look at the energy shifts
introduced by a hypothetical non-Newtonian potential and we
investigate the constraints that FORCA-G could impose on the
strength of this deviation. Finally, in section \ref{Sec:6} we
discuss our results.

\section{The physical system}\label{Sec:2}

In this section we are going to describe the main features of our
physical system and the hamiltonian formalism used to investigate
the interaction between atom and electromagnetic field. Let us
consider a two-level atom trapped in an optical standing wave
produced by the reflection of a laser having wavelength
$\lambda_l=\frac{2\pi}{k_l}$ on a surface located at $z=0$. In the
configuration we are considering the optical trap has a vertical
orientation, so that we have to take into account the earth's
gravitation field acting on the atom. The complete hamiltonian can
be written under the form
\begin{equation}\begin{split}
H&=H_0+H_\text{int}=H_\text{f}+H_\text{at}+H_{\footnotesize\text{WS}}+H_\text{int}\\
H_\text{f}&=\sumpk\,\hbar\omega\,\acpk\apk\\
H_\text{at}&=\hbar\omega_0\ket{e}\bra{e}\\
H_{\footnotesize\text{WS}}&=\frac{p^2}{2m}+mgz+\frac{U}{2}\bigl(1-\cos(2k_lz)\bigr)\\
H_\text{int}&=-\bbm[\mu]\cdot\mathbf{\mathcal{E}}(\mathbf{r}).\end{split}\label{Htot}\end{equation}
The complete Hamiltonian is written as a sum of a term $H_0$
describing the free evolution of the atomic and field degrees of
freedom. In particular, $H_\text{f}$ is the Hamiltonian of the
quantum electromagnetic field, described by a set of modes
$(p,\mathbf{k},k_z)$: here $p$ is the polarization index, taking
the values $p=1,2$ corresponding to TE and TM polarization
respectively, while $\mathbf{k}$ and $k_z$ are the transverse and
longitudinal components of the wavevector. We associate to each
single mode a frequency $\omega=c\sqrt{k^2+k_z^2}$, as well as
annihilation and a creation operators $\apk$ and $\acpk$. An
eigenstate of the field Hamiltonian is thus specified by giving a
set of photon occupation numbers $\ket{\{n_p(\mathbf{k},k_z)\}}$
for each mode of the field. The vacuum state of the field, with
zero photons in each mode, will be noted with
$\ket{0_p(\mathbf{k},k_z)}$. In our formalism the expression of
the electric field is the following
\begin{equation}\begin{split}\mathbf{\mathcal{E}}(\mathbf{r})&=\frac{i}{\pi}\sumpk\sqrt{\frac{\hbar\omega}{4\pi\epsilon_0}}\\
&\,\times\bigl(e^{i\mathbf{k}\cdot\mathbf{r}_\perp}\fpkz\apk-\text{h.c.}\bigr)\end{split}\label{EF}\end{equation}
where we have introduced the transverse coordinate
$\mathbf{r}_\perp=(x,y)$ and the mode functions $\fpkz$
characterizing the boundary conditions imposed on the field. Under
the assumption of a perfectly conducting mirror in $z=0$ these
functions take a very simple expression \cite{BartonJPhysB74}
\begin{equation}\begin{split}\mathbf{f}_1(\mathbf{k},k_z,z)&=\hat{\mathbf{k}}\times\hat{\mathbf{z}}\sin(k_zz)\\
\mathbf{f}_2(\mathbf{k},k_z,z)&=\hat{\mathbf{k}}\frac{ick_z}{\omega}\sin(k_zz)-\hat{\mathbf{z}}\frac{ck}{\omega}\cos(k_zz)\end{split}\end{equation}
where $\hat{\mathbf{k}}=\mathbf{k}/k$ and
$\hat{\mathbf{z}}=(0,0,1)$. $H_\text{at}$ is the internal
Hamiltonian of our two level atom having ground state $\ket{g}$
and excited state $\ket{e}$ separated by a transition frequency
$\omega_0$. While $H_\text{at}$ is associated to the internal
atomic degrees of freedom, the term $H_{\footnotesize\text{WS}}$
accounts for the external atomic dynamics. As a consequence, it
contains the kinetic energy ($p$ being the canonical momentum
associated to $z$), as well as both the gravitational potential
(treated here in first approximation as a linear term), where $m$
is the atomic mass and $g$ is the acceleration of the Earth's
gravity, and the classical description of the stationary optical
trap, having depth $U$. We treat here only the $z$-dependent terms
of the Hamiltonian since the degrees of freedom $x$ and $y$, even
in presence of a transverse trapping mechanism, are decoupled from
the longitudinal dynamics. For simplicity, we shall take as a unit
of energy the photon recoil energy $E_r$ given by
$E_r=\frac{\hbar^2k_l^2}{2m}$. As far as the atomic position is
concerned, it will be expressed in units of the periodicity of the
trap $\frac{\lambda_l}{2}$. For all numerical examples in this
paper we will use the experimental configuration chosen for
FORCA-G: $E_r=5.37\times10^{-30}\,$J
($\frac{E_r}{h}=8.11\times10^3\,$Hz) and
$\frac{\lambda_l}{2}=266\,$nm.

The interaction between the atom and the quantum electromagnetic
field is written here in the well-known multipolar coupling in
dipole approximation \cite{PowerPhilTransRoySocA59}, where
$\bbm[\mu]=q\bbm[\rho]$ ($q$ being the electron's charge and
$\bbm[\rho]$ the internal atomic coordinate) is the quantum
operator associated to the atomic electric dipole moment and the
electric field is calculated in the atomic position $\rv$. It is
important to observe that, since $\bbm[\mu]$ clearly operates only
on the atomic internal states, this interaction term is the only
one coupling atomic (both internal and external) and field degrees
of freedom. As a consequence, the ground state of the free
Hamiltonian $H_0$ is simply given by the tensor product of the
vacuum field state $\ket{0_p(\mathbf{k},k_z)}$, the atomic state
$\ket{g}$ and the ground state of $H_{\footnotesize\text{WS}}$. In
the picture of atomic dressing \cite{CPP95}, the ground state of
$H_0$ is the bare ground state, and the inclusion of
$H_\text{int}$ will produce a new ground state of the complete
system, referred to as dressed ground state, mixing all the
degrees of freedom. We are going to tackle the calculation of the
ground state of $H_{\footnotesize\text{WS}}$ in the next section,
whereas the atom-field interaction will be treated in section
\ref{Sec:4}.

\section{Modified Wannier-Stark states}\label{Sec:3}

\subsection{Ordinary Wannier-Stark states}\label{Ordinary}

In solid state physics, it is well-known that the solution of the
time-independent Schr\"{o}dinger equation describing a quantum
particle in a periodic potential leads to the so-called Bloch
states \cite{BlochZPhys29,Ashcroft}. Due to the periodicity of the
system, these states are completely delocalized in space
coordinate and the spectrum energy is composed of bands of
permitted energies, each band being labeled with an index
$b=1,2,\dots$. The addition of a linear potential (whose role is
in our case played by gravity) to the trap produces localization
of the states: these states are usually labeled as Wannier-Stark
states (see e.g. \cite{WannierPhysRev60,GluckEurJPhysD98}). We
will now describe their main features. For each Bloch band $b$, a
discrete quantum number $n$ is introduced, taking the values
$n=0,\pm1,\pm2,\dots$. The state $\ket{n,b}^\text{\tiny{(WS)}}$
is, in coordinate representation, approximately centered in the
$n$-th well of the optical trap, and the energy of this state is
in first approximation given by
\begin{equation}\label{EWS}E^\text{\tiny{(WS)}}_{n,b}=\bar{E}^\text{\tiny{(WS)}}_b+n\frac{mg\lambda_l}{2E_r}\end{equation}
with $\bar{E}^\text{\tiny{(WS)}}_b$ the average of the $b$-th
Bloch band \cite{NiuPRL96,GluckPhysRep02}. As a result of the
quasi-periodicity of the system (i.e. of the linearity of the
gravitational potential modifying the periodic trap) two states
$\ket{n,b}^\text{\tiny{(WS)}}$ and $\ket{p,b}^\text{\tiny{(WS)}}$
belonging to the same band $b$ are shifted, in coordinate
representation, by $n-p$ wells. At the same time their energies
differ, in accordance with eq. \eqref{EWS}, by $n-p$ times
$\delta_g=\frac{mg\lambda_l}{2E_r}$. Then, the problem of
Wannier-Stark states is solved once we know, for each band $b$,
the average Bloch-band energy $\bar{E}^\text{\tiny{(WS)}}_b$ and
the eigenfunction centered in a given well. The Wannier-Stark
states can be calculated using, for example, the numerical
approach of \cite{GluckEurJPhysD98}.

\begin{figure}[h]\centering
\includegraphics[height=5cm]{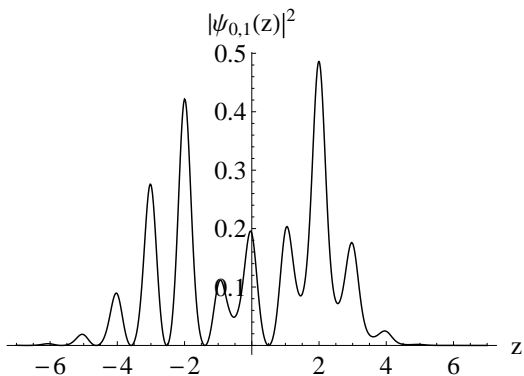}
\caption{Coordinate representation of the state $\ket{0,1}$
belonging to the first Bloch band and centered in the zeroth well
for $U=3$.}\label{U3WS}\end{figure}

\begin{figure}[h]\centering
\includegraphics[height=5cm]{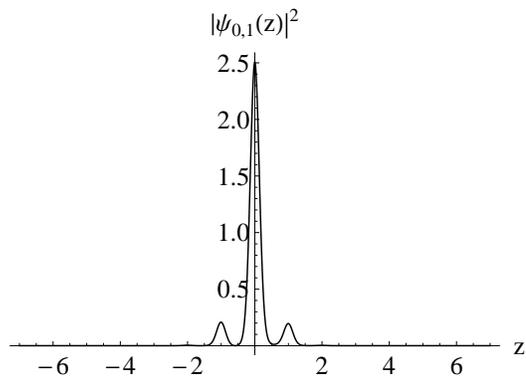}
\caption{Coordinate representation of the state $\ket{0,1}$
belonging to the first Bloch band and centered in the zeroth well
for $U=10$.}\label{U10WS}\end{figure}

In figures \ref{U3WS} and \ref{U10WS} we give the Wannier-Stark
states $\ket{0,1}^\text{\tiny{(WS)}}$ for two different values of
the potential depth $U=3,10$ (in units of $E_r$).
The figures show that, as
expected, a deeper well produces a more localized state of the
particle.

\subsection{Wannier-Stark states in proximity of a surface}

In the context of our problem, the presence of a surface at $z=0$
plays two roles. On one hand, it induces a modification of the
Wannier-Stark states by imposing a boundary condition on the
eigenvalue problem. On the other hand, the quantum
electrodynamical interaction between the atom and this surface
must be taken into account, as we will describe in section
\ref{Sec:4}.

The surface at $z=0$ breaks the quasi-periodicity of the system.
The potential modifying the optical trap is no longer linear,
since it must be considered as the gravitational linear potential
for $z>0$ and an infinite potential barrier for $z\leq0$,
describing the impossibility of the particle to penetrate into the
mirror. We will refer to the eigenstates of this new physical
system as the \emph{modified} Wannier-Stark states. From now on we
are going to deal only with these new states: the state of the
$b$-th Bloch band centered in the $n$-th well will be noted with
$\ket{n,b}$ (and correspondingly $\psi_{n,b}(z)$).

We have solved the problem of modified Wannier-Stark states
numerically, using a finite-difference method. The first step of
our approach consists in considering a unidimensional box
$0<z<z_f$ and imposing that the wavefunction vanishes at the
borders. As for $z=0$, this corresponds to a real physical
boundary condition, whereas the condition $\psi(z_f)=0$ is purely
numerical. Naturally, the acceptability of the solutions will
depend on their rate of decay toward 0 for $z\to z_f$. The next
step is the discretization of the interval $[0,z_f]$ using a set
of $N+2$ mesh points $z_i$ with $z_0=0,z_1,\dots,z_{N+1}=z_f$
(giving $\delta z=\frac{z_f}{N+1}$ for equally spaced mesh
points).

Using this approach, the problem is reduced to an eigenvalue
problem of a tridiagonal symmetric matrix. The solution of such a
problem can be efficiently worked out using the numerical approach
first introduced in \cite{PetersTheComputerJournal69} as well as a
standard QL algorithm \cite{Press95}. In order to check the
robustness of our numerical results, we have also checked their
coherence with a finite-element method
\cite{AbrashkevichCompPhysComm95,SukumarIntJNumerMethEngng09}.

Choosing a large enough numerical box, taking for example $z_f=30$
(we recall here that $z$ is measured in units of trap periods
$\frac{\lambda_l}{2}$), we have verified that the modified
Wannier-Stark states centered in a well far from the surface
(approximately starting from $n=10$) have the same shape as the
functions shown in section \ref{Ordinary}: this reflects the fact
that far from the surface the quasi-periodicity of the system is
reestablished. Moreover, in this region we find that the energy
difference between two successive states equals the expected
quantity $\delta_g$ defined before: starting from $n=10$, the
differences equal $\delta_g$ with a relative precision better than
$10^{-4}$. This can be seen in table \ref{TableE3}, where we show
the results obtained for the first ten energy levels with $U=3$:
in this table we give the energy levels $E_n$, as well as the
differences $\delta E_n=E_{n+1}-E_n$. In this configuration we
have $\delta_g=0.070068$.
\begin{widetext}\begin{center}
\begin{figure}[h!]\centering
\includegraphics[height=5cm]{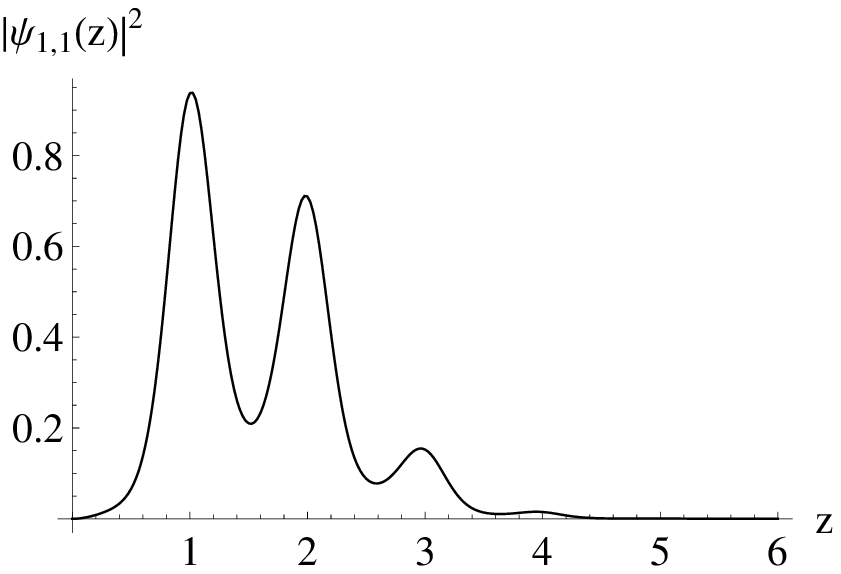}\hspace{1cm}\includegraphics[height=5cm]{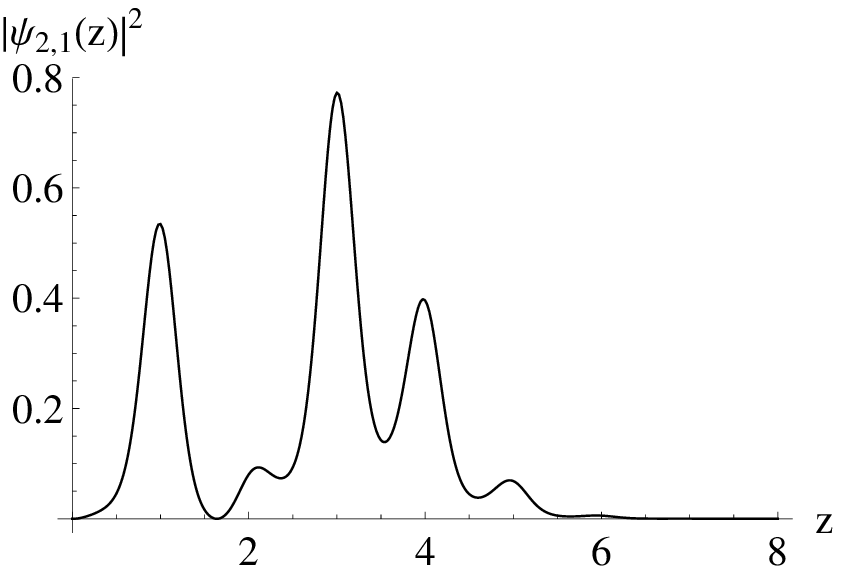}\\
\includegraphics[height=5cm]{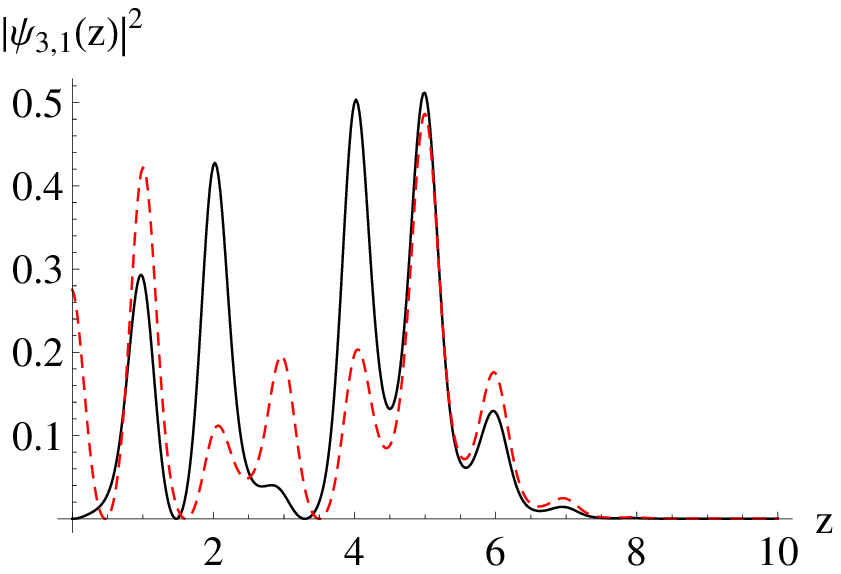}\hspace{1cm}\includegraphics[height=5cm]{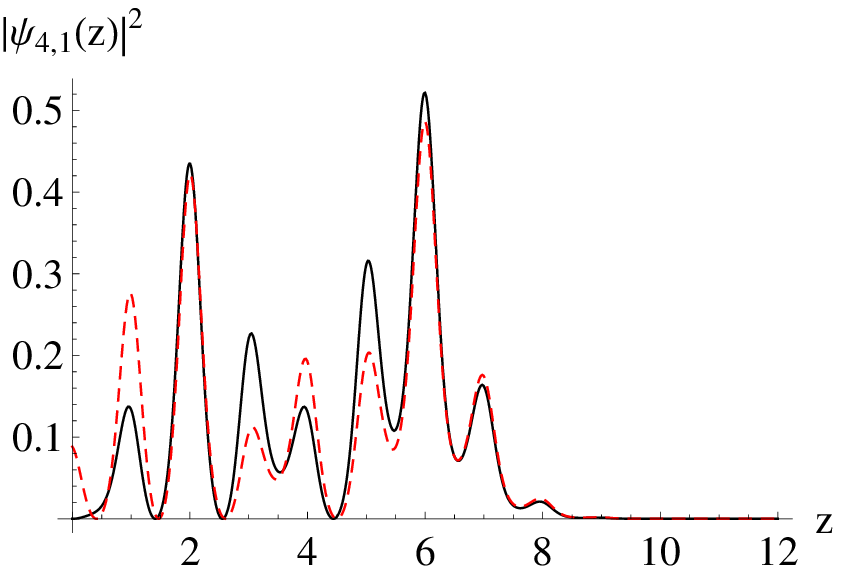}
\caption{(Color online) Density probability of modified
Wannier-Stark states $\psi_{n,1}(z)$ for $n=1,2,3,4$ and $U=3$.
The last two functions (black, solid line) are compared to the
corresponding standard Wannier-Stark state (red, dashed
line).}\label{Psi}\end{figure}\end{center}\end{widetext}Table
\ref{TableE3} shows only values of the energies belonging to the
first Bloch band ($b=1$). We have checked that, increasing the
value of $N$ above $10^6$, the last digit reported in table
\ref{TableE3} remains constant, corresponding to a relative
precision of approximately $10^{-4}$. Actually, as a result of our
numerical method we also found the eigenvalues and corresponding
states associated to higher bands. Nevertheless, we shall discuss
only the first-band states since the ones belonging to higher
bands are much less relevant for experimental purposes: as a
matter of fact, higher bands are not efficiently trapped in the
experiment (the average energy of the second band is around $4E_r$
for a trap depth of $3E_r$).
\begin{center}\begin{table}[h]\begin{tabular}{|c|c|c|c|}
\hline & & & \vspace{-.3cm}\\
$n$ & $E_n$ & $\delta E_n (\times10^{-2})$ & $\delta E_n (\times10^2 \text{Hz})$\\
\hline\hline 1 & 1.4028 & 12.302 & 9.9788\\2 & 1.5258 & 9.8043 &
7.9525\\3 & 1.6239 & 8.4432 & 6.8485\\4 & 1.7083 & 7.6206 &
6.1812\\5 & 1.7845 & 7.2026 & 5.8422\\6 & 1.8566 & 7.0518 &
5.7199\\7 & 1.9271 & 7.0146 & 5.6897\\8 & 1.9972 & 7.0079 &
5.6843\\9 & 2.0673 & 7.0070 & 5.6835\\10 & 2.1374 & 7.0068 &
5.6834\\11 & 2.2074 & 7.0068 & 5.6834\\12 & 2.2775 & 7.0068 &
5.6834\\13 & 2.3476 & 7.0068 &
5.6834\\\hline\end{tabular}\caption{First ten values of the
modified Wannier-Stark spectrum for $U=3$. These values have been
obtained on an interval $[0,30]$. The first two columns are in
units of $E_r$, the third one is in
Hz.}\label{TableE3}\end{table}\end{center}As far as the states in
proximity of the plate are concerned, they are strongly modified
by the boundary condition, and the same property holds for their
energies. We show, in figure \ref{Psi} the first four
eigenfunctions in presence of the surface. For the sake of
comparison, the third and fourth wavefunctions are superposed to
the standard Wannier-Stark solutions centered in the corresponding
well. It is important to stress that the ordinary Wannier-Stark
functions of wells $n=3,4$ are plotted only to show that the shape
of the modified ones tends towards the standard solution: however,
the fact that the ordinary functions for these wells are different
from zero for $z\leq0$ makes strictly speaking no sense for our
physical system.

\begin{figure}[h]\centering
\includegraphics[height=5.5cm]{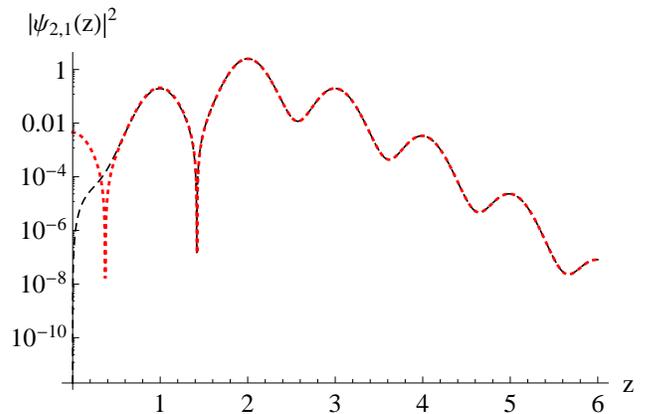}
\caption{(Color online) Density probability of modified
Wannier-Stark states $\psi_{2,1}(z)$ (black, dashed line) for
$U=10$ compared to the corresponding standard Wannier-Stark state
(red, dotted).}\label{PsiU10}\end{figure}

In order to discuss the influence of the depth of the wells, we
will conclude this section giving the results obtained for $U=10$.
In this case, since the ordinary Wannier-Stark states are much
more localized in each well, we expect the influence of the
surface to be evident on a smaller range of distances. This can be
seen directly from table \ref{TableE10}, where the energy
differences converge more rapidly to $\delta_g$.
\begin{center}\begin{table}[h]\begin{tabular}{|c|c|c|c|}
\hline & & & \vspace{-.3cm}\\
$n$ & $E_n$ & $\delta E_n (\times10^{-2})$ & $\delta E_n (\times10^2 \text{Hz})$\\
\hline\hline 1 & 2.9496 & 7.5127 & 6.0938\\2 & 3.0247 & 7.0276 &
5.7003\\3 & 3.0950 & 7.0072 & 5.6837\\4 & 3.1651 & 7.0068 &
5.6834\\5 & 3.2352 & 7.0068 & 5.6834\\6 & 3.3052 & 7.0068 &
5.6834\\7 & 3.3753 & 7.0068 & 5.6834\\8 & 3.4454 & 7.0068 &
5.6834\\9 & 3.5154 & 7.0068 & 5.6834\\10 & 3.5855 & 7.0068 &
5.6834\\\hline\end{tabular}\caption{First ten values of the
modified Wannier-Stark spectrum for $U=10$. Same parameters as for
table \ref{TableE3}.}\label{TableE10}\end{table}\end{center}
Moreover, from figure \ref{PsiU10} we see that the state
$\psi_{2,1}(z)$ shows already a remarkable accordance with the
corresponding unmodified Wannier-Stark state in the interval
$[0.6,6]$, where the probability of finding the atom is
approximately $0.9997$.

\section{Casimir-Polder interaction}\label{Sec:4}

\subsection{Standard Casimir-Polder calculations}\label{StandardCP}

The presence of the surface does not only play the role of
imposing a boundary condition on the Wannier-Stark wavefunctions.
In fact, since it modifies the structure of the modes of the
quantum electromagnetic field, it is source of an attractive force
between the atom and the plate. This is a particular case of a
general phenomenon usually called Casimir effect for two
macroscopic bodies and Casimir-Polder force when it involves one
or more atoms near a surface (for a general review see e.g.
\cite{Milonni}). This phenomenon was first pointed out by Casimir
in 1948 for two parallel perfectly conducting plates
\cite{CasimirProcKonNederlAkadWet48} and in the same year by
Casimir and Polder for atom-surface and atom-atom systems
\cite{CasimirPhysRev48}.

The Casimir-Polder force between an atom and a mirror has been
measured quite recently using several different techniques:
deflection of atomic beams \cite{SukenikPRL93}, reflection of cold
atoms \cite{LandraginPRL96,ShimizuPRL01,DruzhininaPRL03}. In the
past few years, Bose-Einstein condensates proved to be efficient
probes of this effect, both by means of reflection techniques
\cite{PasquiniPRL04,PasquiniPRL06} and observing center-of-mass
oscillations of the condensate
\cite{AntezzaPRA04,HarberPRA05,AntezzaPRL06,ObrechtPRL07}. The
FORCA-G experiment aims at achieving a percent precision in the
measurement of the force thanks to the combination of cold atoms
and interferometric techniques.

From a theoretical point of view, the force is usually obtained
from an interaction energy which results from a time-independent
perturbative calculation on the matter-field hamiltonian
interaction term \cite{CPP95,ScheelActaPhysSlov08}. In this kind
of approach, the position of the atom is usually treated as a
fixed parameter and not as a quantum operator. As a consequence,
in order to deduce the Casimir-Polder interaction energy between
an atom and a perfectly conducting plate, we must neglect the term
$H_{\footnotesize\text{WS}}$ in the Hamiltonian of the system
\eqref{Htot} and use eq. \eqref{EF} for the electric field.
Choosing the bare ground state $\ket{0_p(\mathbf{k},k_z)}\ket{g}$
as the unperturbed configuration, the first-order perturbative
correction on interaction term $H_\text{int}$ is zero, since the
atomic electric dipole moment operator is an odd operator and the
annihilation and creation operators appearing in the electric
field do not connect states with the same number of photons.
Moving to second-order, we obtain the $z$-dependent potential
energy
\begin{equation}\label{VTZ}V^{(2)}_\text{CP}(z)=-\sumpk\frac{\bigl|\Apkz\bigr|^2}{\hbar(\omega+\omega_0)}.\end{equation}
In this expression we have defined
\begin{equation}\begin{split}\Apkz&=\bra{0_p(\mathbf{k},k_z)}\bra{g}H_\text{int}\ket{1_p(\mathbf{k},k_z)}\ket{e}\\
&=-\frac{i}{\pi}\sqrt{\frac{\hbar\omega}{4\pi\epsilon_0}}e^{i\mathbf{k}\cdot\mathbf{r}_\perp}\bbm[\mu]\cdot\fpkz\end{split}\end{equation}
and we sum over all the possible intermediate states
$\ket{1_p(\mathbf{k},k_z)}\ket{e}$ having one photon in the mode
$(p,\mathbf{k},k_z)$ and the atom in its excited internal state
$\ket{e}$. Finally, the superscripts $(2)$ and $(1)$ refer to the
order with respect to the electric charge contained in
$\bbm[\mu]$.

This result holds for a perfectly conducting surface and at zero
temperature. However, the generalization to more realistic
configurations including the finite conductivity of the plate as
well as a temperature $T>0$ is not straightforward in a
perturbative approach. This can be worked out using for example
the scattering method \cite{LambrechtNewJPhys06,MessinaPRA09} or
the Green-function formalism (see \cite{WyliePRA84,BuhmannPRA05}
and references therein). The resulting potential can be put under
the form \cite{FootnoteCasimir}
\begin{equation}\label{Vz}\begin{split}V_\text{CP}^{(2)}(z)&=\frac{2k_BT}{c^2}\sum_{n=0}^{+\infty}{}'\xi_n^2\frac{\alpha(i\xi_n)}{4\pi\epsilon_0}\int_0^{+\infty}dk\,\frac{ke^{-2K_nz}}{2K_n}\\
&\,\times\Bigl[r_\TE(k,i\xi_n)-\Bigl(1+\frac{2c^2k^2}{\xi_n^2}\Bigr)r_\TM(k,i\xi_n)\Bigr].\\\end{split}\end{equation}
where $\xi_n=\frac{2\pi nk_BT}{\hbar}$ is the $n$-th Matsubara
frequency and the prime on the Matsubara sum indicates that the
$n=0$ term is to be taken with half weight. Moreover we have
defined $K_n=\sqrt{\frac{\xi_n^2}{c^2}+k^2}$ and the
$r_p(k,\omega)$ are the well-known Fresnel coefficients for a
planar surface. Finally $\alpha(\omega)$ is the ground-state
atomic polarizability, which for a multilevel atom takes the form
\cite{FootnotePolarizability}
\begin{equation}\label{Polar}\alpha(\omega)=\frac{2}{3}\sum_n\frac{E_{n0}\mu_{n0}^2}{E_{n0}^2-\hbar^2\omega^2}\end{equation}
where $E_{n0}=E_n-E_0$ is the difference between the energies of
the $n$-th atomic level (starting from the first excited state)
and of ground state, whereas $\mu_{n0}$ is the matrix element of
the electric dipole operator between the same couple of states.
Clearly, the conductive properties of the surface material are
included in the Fresnel coefficients through the electric
permittivity and magnetic susceptibility $\epsilon(\omega)$ and
$\mu(\omega)$ respectively. We conclude this section giving the
expression of the Casimir-Polder potential for an atom in front of
a real surface at zero temperature
\begin{equation}\label{Vz0}\begin{split}V_\text{CP}^{(2)}(z)&=\frac{\hbar}{\pi c^2}\int_0^{+\infty}d\xi\,\xi^2\frac{\alpha(i\xi)}{4\pi\epsilon_0}\int_0^{+\infty}dk\,\frac{ke^{-2Kz}}{2K}\\
&\,\times\Bigl[r_\TE(k,i\xi)-\Bigl(1+\frac{2c^2k^2}{\xi^2}\Bigr)r_\TM(k,i\xi)\Bigr]\\\end{split}\end{equation}
where $K=\sqrt{\frac{\xi^2}{c^2}+k^2}$ and the sum over the
Matsubara frequencies is replaced by an integral.

\subsection{Perturbation of modified Wannier-Stark}

\begin{figure}[h]\centering
\includegraphics[height=5cm]{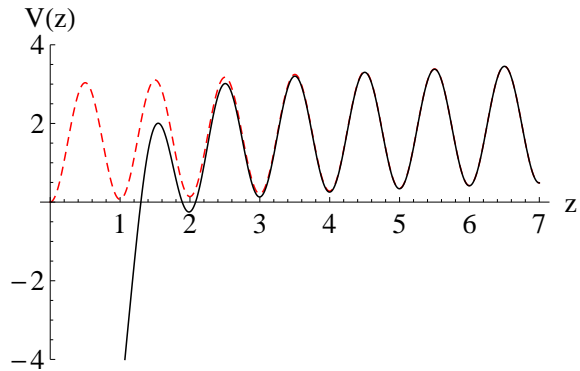}
\caption{(Color online) Sum of Wannier-Stark (for $U=3$) and
Casimir-Polder potentials (in black, solid line) compared to
Wannier-Stark potential alone (in red, dashed
line).}\label{Potential}\end{figure}

In the last section we have given the Casimir-Polder potential for
an atom having polarizability $\alpha(\omega)$ in front of an
arbitrary planar surface and at temperature $T$. It could be
natural to think that this $z$-dependent potential should be added
to the Wannier-Stark Hamiltonian $H_{\footnotesize\text{WS}}$ in
\eqref{Htot} to obtain a new time-independent problem. So one
could obtain a new set of energies and wavefunctions taking also
into account the quantum electrodynamical part of the problem. In
figure \ref{Potential} we plot this new \emph{complete} potential
(sum of \eqref{VTZ} and the ordinary Wannier-Stark potential) for
a Rubidium atom in front of a perfectly conducting surface at zero
temperature.
This plot clearly shows that
the Casimir-Polder interaction modifies the optical trap on a
limited range. In particular, in our case the first well does no
longer exist, the second and the third are slightly modified, and
starting from the fourth the trap is practically unperturbed.

Nevertheless, the simple addition of the Casimir-Polder
$z$-dependent potential to the external hamiltonian term
$H_\text{WS}$ is strictly speaking incorrect. As a matter of fact,
the potential \eqref{VTZ} (as well as \eqref{Vz} and \eqref{Vz0})
has been derived using several hypotheses. First, it arises from a
perturbative treatment of the interaction term $H_\text{int}$ on
the Hamiltonian $H_\text{at}+H_\text{f}$. Moreover, in this
calculation the atomic position $z$ is treated as a fixed
parameter. This is clearly incoherent with the fact that for our
complete Hamiltonian \eqref{Htot} $z$ is a dynamic variable.

These arguments suggest that we should reconsider the calculation
of the Casimir-Polder potential using our perturbative approach
including the Wannier-Stark Hamiltonian
$H_{\footnotesize\text{WS}}$. The perturbative term is still
$H_\text{int}$, but now the atomic coordinate $z$ has to be
treated as a quantum operator as well. So, we are able to
introduce a new unperturbed state for each well $n$ of the first
Bloch band having the form
\begin{equation}\ket{\psi^{(0)}_{n,1}}=\ket{0_p(\mathbf{k},k_z)}\ket{g}\ket{n,1}.\end{equation}
As we found for ordinary Casimir-Polder calculation, the
leading-order correction to the energies is the second, and the
corrections takes the new form
\begin{equation}\label{Pertu}\begin{split}\Delta
E^{(2)}_{n,1}&=-\sumpk\\
&\,\times\sumpb\frac{\bigl|\bra{\psi^{(0)}_{n,1}}H_\text{int}\ket{1_p(\mathbf{k},k_z)}\ket{e}\ket{p,b}\bigr|^2}{E^{(0)}_{p,b}-E^{(0)}_{n,1}+\hbar(\omega+\omega_0)}\end{split}\end{equation}
where now the intermediate state contains the modified
Wannier-Stark state $\ket{p,b}$. We notice that the difference
between two Wannier-Stark energies appearing in the denominator is
$p-n$ times approximately $0.07E_r$ if the state $\ket{p,b}$
belongs to the first band ($b=1$). We now point out that the
recoil energy for a Rubidium atom having $m=1.44\cdot10^{-25}\,$kg
trapped in a periodic potential having $\lambda_l=532\,$nm is of
the order of $10^{-11}\,$eV. On the other hand, the atomic
transition energy $\hbar\omega_0$ is of the order of the eV. At
the same time, the numerator in \eqref{Pertu} involves an integral
over the $z$ coordinate containing the product of the
wavefunctions associated to the two states. This product becomes
negligibly small for $|p-n|\gtrsim7$, whilst the energy difference
$E^{(0)}_{p,1}-E^{(0)}_{n,1}$ is still orders of magnitude smaller
than $\hbar\omega_0$. As a consequence, the Wannier-Stark energy
difference in the denominator can be always safely neglected with
respect to $\hbar\omega_0$ for the intermediate states having
$b=1$. As far as the higher bands are concerned, it is possible to
see that the same superposition integral decays to zero due to the
delocalization of the modified Wannier-Stark states. Furthermore,
in the case of higher bands, the energy difference
$E^{(0)}_{p,b}-E^{(0)}_{n,1}$ is still $100E_r$ (and then still
negligible with respect to $\hbar\omega_0$) for $b=10$. This
reasoning enables us to use the closure relation on the
$\ket{n,b}$ states and obtain
\begin{equation}\begin{split}\Delta &E^{(2)}_{n,1}=\\
&\,\bra{n,1}\Bigl(-\sumpk\frac{\bigl|\Apkz\bigr|^2}{\hbar(\omega+\omega_0)}\Bigl)\ket{n,1}.\end{split}\end{equation}
The expression in parentheses coincides with the second-order
perturbative calculation on the atom-field ground state described
in section \ref{StandardCP}. It is thus evident that the
correction we are looking for equals the average on the
Wannier-Stark state $\ket{n,1}$ of the known Casimir-Polder
potential $V_\text{CP}^{(2)}(z)$. This can be then expressed as
follows
\begin{equation}\label{Correc}\Delta E^{(2)}_{n,1}=\int_0^{+\infty}dz\,\bigl|\psi^{(0)}_{n,1}(z)\bigr|^2V^{(2)}_\text{CP}(z).\end{equation}
This expression has been obtained in the context of a perturbative
treatment for a perfectly conducting surface at zero temperature.
Nevertheless, the reasoning which led us from the general
expression \eqref{Pertu} to the simple average value
\eqref{Correc} does not depend on the details of the calculation
of the $V^{(2)}_\text{CP}(z)$ itself. As a consequence, it is
reasonable to assume that the average value \eqref{Correc} can be
also used with the more general expressions of the interaction
energy \eqref{Vz} or \eqref{Vz0}.

The behavior of the integrand function around $z=0$ must be
treated with care: indeed, the Casimir-Polder potential diverges
for $z\to0$. In particular, it is well-known that for distances
much smaller than the typical atomic transition wavelength (van
der Waals regime) the interaction potential is
temperature-independent and its expression reads
\cite{AntezzaPRA04}\begin{equation}V_{\text{CP;vdW}}^{(2)}(z)=-\frac{\hbar}{4\pi
z^3}\int_0^{+\infty}d\xi\,\frac{\alpha(i\xi)}{4\pi\epsilon_0}\frac{\epsilon(i\xi)-1}{\epsilon(i\xi)+1}\end{equation}
$\epsilon(\omega)$ being the electric permittivity of the surface
material. As for the atomic wavefunction, we have numerically
verified that, for any allowed value of $n$, it tends to zero
linearly for $z\to0$. As a consequence, the integrand function
behaves like $z^{-1}$ around the origin, implying a divergent
energy correction \eqref{Correc} for any $n$. We will develop in
the next section an effective description of the atom to
regularize this quantity.

\subsection{Regularization of the correction}

The potential $V_\text{CP}^{(2)}(z)$ represents a particular case
of singular potentials since it diverges around the origin faster
than $z^{-2}$. The treatment of such potentials has been discussed
since the pioneering work of Case \cite{CasePhysRev50} (for more
details see e.g. \cite{FrankRevModPhys71}). It can be shown from
first principles \cite{Landau77} that these potentials describe an
unphysical situation in proximity of the origin. The solution of
the time-independent Schr\"{o}dinger equation requires in these
cases a more detailed knowledge of the short-distance physics of
the problem.

In the case of atom-surface interaction, the $z^{-3}$ behavior of
the potential is an artefact of treating the atom as a point-like
source. This statement is supported by the calculation of the
Casimir potential between a sphere of radius $R$ and a wall
\cite{MaiaNetoPRA08,CanaguierPRL09,CanaguierArxiv}. The potential
energy associated with this geometrical configuration shows a
$z^{-4}$ long-distance behavior (equivalent to the long-distance
Casimir-Polder atom-surface interaction), an intermediate $z^{-3}$
regime and a transition toward a $z^{-1}$ behavior when
approaching $z=0$. This property holds for any value of the radius
$R$ of the sphere: nevertheless, the characteristic distance at
which the transition occurs is, as physically predictable, of the
order of the radius.

Inspired by \cite{CompagnoEurophysLett88}, we take into account
the finite size of the atom by replacing it with a probability
density distribution $\rho(\mathbf{r}')$. In accordance with
\cite{CompagnoEurophysLett88}, we make the further assumption that
the function $\rho(\mathbf{r}')$ is different from zero within a
finite volume. Moreover, in our numerical applications, we take
this volume to be a sphere of radius $R$, discussing also the
dependence of the results on $R$. We assume that the atom has
coordinate $(0,0,z)$. We stress here that the atomic coordinate
$z$ is taken at the point of the sphere nearest to the surface: as
a consequence the effective sphere representing the atom is
centered in $(0,0,z+R)$. As far as the probability density
distribution is concerned, we will consider the cases of a
constant function $\rho_1(\mathbf{r}')=N_1$ and of a spherically
symmetric parabolic distribution
\begin{equation}\rho_2(\mathbf{r}')=N_2\Bigl[R^2-x'^2-y'^2-(z'-z-R)^2\Bigr].\end{equation}
For both probability distributions the variable $\mathbf{r}'$ is
expressed in the same frame of reference as for the atomic
coordinate. The factors $N_1$ and $N_2$ are to be deduced from the
normalization condition
\begin{equation}\int_\Omega d^3\mathbf{r}'\,\rho(\mathbf{r}')=1\end{equation}
being $\Omega$ the spherical atomic volume. Our hypothesis leads
to a new regularized expression of the atom-surface potential,
given by the average with respect to $\rho(\mathbf{r}')$ of the
standard Casimir-Polder potential
\begin{equation}\label{Vreg}V_{\text{CP;reg}}^{(2)}(z)=\int_\Omega d^3\mathbf{r'}\,\rho(\mathbf{r}')V_{\text{CP}}^{(2)}(z')\end{equation}
where the $z$-dependence of the new potential is implicitly
contained in the probability density distribution
$\rho(\mathbf{r}')$ and the integration volume $\Omega$.
Substituting \eqref{Vreg} into \eqref{Correc} then provides the
regularized energies of our system.

\begin{figure}[h]\centering
\includegraphics[height=5cm]{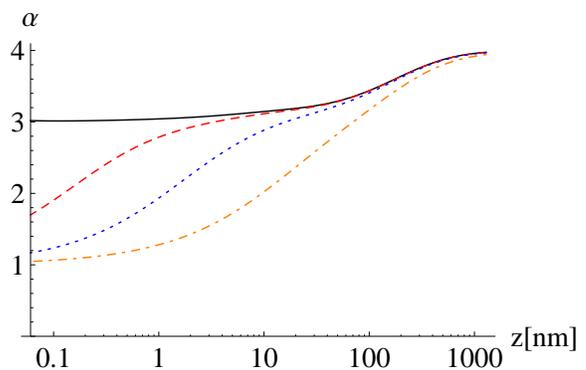}
\caption{(Color online) Exponent $\alpha$ defined in
\eqref{Exponent} for the standard Casimir-Polder potential
\eqref{Vz0} (black, solid line), and the regularized potential
\eqref{Vreg} for radii 0.1 nm (red, dashed line), 1 nm (blue,
dotted line) and 10 nm (orange, dotted-dashed
line).}\label{Exp}\end{figure}

Let us now analyze the behavior of the regularized potential
\eqref{Vreg} in proximity of the surface. Assuming that it has a
form
\begin{equation}V_{\text{CP;reg}}^{(2)}(z)=\frac{A}{z^\alpha}\end{equation}
the exponent $\alpha$ has the form
\begin{equation}\label{Exponent}\alpha=-z\frac{\partial_zV_{\text{CP;reg}}^{(2)}(z)}{\partial z}\frac{1}{V_{\text{CP;reg}}^{(2)}(z)}.\end{equation}
In figure \ref{Exp} we plot the exponent $\alpha$ as a function of
$z$ for the standard Casimir-Polder potential \eqref{Vz0} and the
regularized one \eqref{Vreg}. Both are calculated in this case for
a Rubidium atom in front of a perfectly conducting surface and at
zero temperature: the data for the dynamical atomic polarizability
of Rubidium were kindly provided by Derevianko et al.
\cite{DereviankoPRL99}. Furthermore, the regularized expression is
calculated for a uniform probability density distribution and
three different radii
$R=100\,\text{pm},1\,\text{nm},10\,\text{nm}$.
In the four cases it is evident that
the transition from $z^{-4}$ to $z^{-3}$ behavior starts around
the first atomic transition wavelength ($\simeq780\,$nm).
Moreover, while for the standard Casimir-Polder calculation the
exponent tends to 3, in all the other cases the finite size of the
atom leads, as anticipated, to a $z^{-1}$ asymptotic dependence.
The figure shows clearly that the length scale of this second
power-law transition is roughly of the order of the atomic size.
We will make use of this regularized potential in the next section
to work out the perturbative calculations on the modified
Wannier-Stark states.

\subsection{Energy corrections}

We are now ready to evaluate the average value \eqref{Correc} of
the potential \eqref{Vreg} on any modified Wannier-Stark state.
Our approach leaves as free parameters the atomic effective radius
$R$ and the probability density distribution $\rho(\mathbf{r}')$.
As far as the radius $R$ is concerned, we first remark that
several non-equivalent definitions of the effective atomic radius
exist in literature. For example Slater gives in
\cite{SlaterJChemPhys64} an empirical value of Rubidium radius
equal to 235 pm with an associated accuracy of 5 pm. On the
contrary, the work \cite{ClementiJChemPhys67} estimates the atomic
radius for Rubidium to be 265 pm. As a consequence, in order to
study the dependence of the results on the value of the radius, we
will consider the two extreme cases $R=200\,$pm and $R=300\,$pm.
As for the probability distribution, we will use the functions
$\rho_1(\mathbf{r}')$ and $\rho_2(\mathbf{r}')$ discussed before.
Furthermore we are going to consider the case of a perfect
conductor for the surface in order to get an insight on the
qualitative features of the energy correction. In table
\ref{TableCorr1} we show the energy corrections to the first
twelve modified Wannier-Stark states obtained choosing two radii
and two probability density distributions. In order to be coherent
with the description of the atom as a sphere, the same
regularization treatment used for the Casimir-Polder interaction
is applied to the other $z$-dependent hamiltonian terms contained
in $H_\text{WS}$. As expected ($R\ll\lambda_l$) this does not
modify our results by more than $10^{-3}$ in relative value.
\begin{center}\begin{table}[h]\begin{tabular}{|c|l|l|l|l|}
\hline
$n$ & 200 pm - $\rho_1$ & 200 pm - $\rho_2$ & 300 pm - $\rho_1$ & 300 pm - $\rho_2$\\
\hline\hline
1 & 2.39[1] & 2.37[1] & 2.19[1] & 2.16[1]\\
2 & 1.76[1] & 1.74[1] & 1.60[1] & 1.58[1]\\
3 & 1.20[1] & 1.18[1] & 1.09[1] & 1.08[1]\\
4 & 6.80 & 6.7147 & 6.18 & 6.09\\
5 & 2.89 & 2.8557 & 2.63 & 2.59\\
6 & 8.71[-1] & 8.60[-1] & 7.91[-1] & 7.80[-1]\\
7 & 1.93[-1] & 1.91[-1] & 1.76[-1] & 1.74[-1]\\
8 & 3.57[-2] & 3.53[-2] & 3.27[-2] & 3.23[-2]\\
9 & 6.76[-3] & 6.70[-3] & 6.33[-3] & 6.27[-3]\\
10 & 1.84[-3] & 1.83[-3] & 1.78[-3] & 1.78[-3]\\
11 & 8.36[-4] & 8.35[-4] & 8.29[-4] & 8.29[-4]\\
12 & 5.10[-4] & 5.10[-4] & 5.09[-4] & 5.09[-4]\\
\hline\end{tabular}\caption{Absolute value (in Hz) of the
Casimir-Polder energy corrections (they are all changed in sign)
to the first twelve modified Wannier-Stark states for $U=3$. The
notation $a[b]$ corresponds to $a\times10^b$. These values are
calculated for a perfectly conducting
surface.}\label{TableCorr1}\end{table}\end{center}It is easy to
see from table \ref{TableCorr1} that, as far as the energy levels
are concerned, a change in the effective radius produces a more
remarkable effect than a change of distribution from the uniform
case $\rho_1(\mathbf{r}')$ to the parabolic $\rho_2(\mathbf{r}')$.
In particular, switching from 200 pm to 300 pm gives a relative
error which is of the order of 10\% on the first wells and then
drops down, whereas the relative correction from
$\rho_1(\mathbf{r}')$ to $\rho_2(\mathbf{r}')$ is at most around
1\%.

\begin{figure}[h]\centering
\includegraphics[height=5cm]{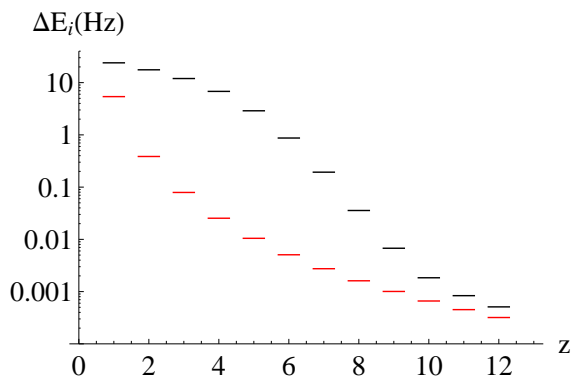}
\caption{(Color online) Absolute value of the Casimir-Polder
energy correction for a uniform distribution and a radius of 200
pm (black upper ticks) compared to the Casimir-Polder potential by
evaluating \eqref{Vz} at the well centre (red lower ticks). The
depth of the trapping potential is $U=3$.}\label{Dere}\end{figure}

It is now instructive to compare one of the set of energy
corrections shown in table \ref{TableCorr1} with the simple
evaluation of the strength of the potential energy \eqref{Vz0} at
the centre of each well \cite{WolfPRA07,DereviankoPRL09}, which
could be used as a first estimation of the energy correction. This
idea works better considering deeper traps or farther from the
surface: for example, we have verified that the value of
$V_\text{CP}^{(2)}(z)$ calculated at $z=1$ and the first energy
correction differ by a factor of approximately $4.4$ for $U=3$
(see figure \ref{Dere}), while this factors drops already to
$1.12$ for $U=20$ and to $1.05$ for $U=80$. We also remark that a
larger value of $U$ or a larger atom-surface distance reduces the
dependence of the results on the choice of both the probability
density distribution and the effective radius. This reasoning
proves that in the case of the experiment FORCA-G the
delocalization of the atom indeed plays a role.

We want to stress here that the validity of our spherical-atom
model used for the regularization of $V_\text{CP}^{(2)}(z)$ still
remains to be tested by experimental measurements. Some more
details, as well as the relationship with the search for
non-Newtonian gravity, will be given in section \ref{Sec:6}.

\section{Deviations from Newtonian gravitation}\label{Sec:5}

\begin{figure}[h]\centering
\includegraphics[height=6.5cm]{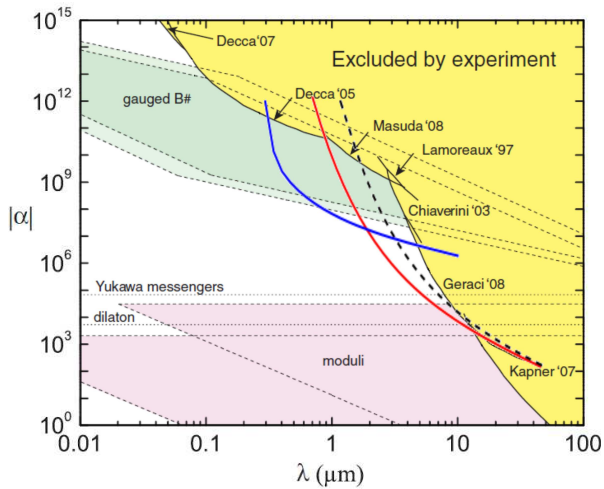}
\caption{(Color online) In yellow are displayed the regions of the
$(\alpha_{\footnotesize\text{Y}},\lambda_{\footnotesize\text{Y}})$
plane excluded by experiments. The figure is taken from
\cite{GeraciPRL10}. The three superposed curves represent the
experimental constraints theoretically calculated for the
experiment FORCA-G. They correspond to the near regime, using a
superposition between wells $n=4$ and $n=6$ (blue solid line,
first from the left), the far regime for $n=40$ (red solid line)
and for $n=70$ (black dashed line).}\label{FigConstr}\end{figure}

Many theories of unification of general relativity and quantum
mechanics predict a modification of the laws of gravity at short
distances. These modifications can be described by the addition of
a new potential to the standard Newtonian one. This correction is
often modelized by a Yukawa-type law so that the complete
gravitational potential between two point-like particles is
written under the form
\begin{equation}\label{YCorr}U_\text{G}(z)=\frac{GMm}{z}\Bigl(1+\alpha_{\footnotesize\text{Y}}e^{-\frac{z}{\lambda_{\footnotesize\text{Y}}}}\Bigr)\end{equation}
where $G$ is the gravitational constant, $m$ and $M$ the masses of
the two particles. In this expression
$\alpha_{\footnotesize\text{Y}}$ and
$\lambda_{\footnotesize\text{Y}}$ are two parameters introduced to
characterize respectively the relative strength of the corrective
potential and its typical range. The experiments aimed at testing
the existence of such a deviation set constraints on the allowed
values of the parameters $\alpha_{\footnotesize\text{Y}}$ and
$\lambda_{\footnotesize\text{Y}}$. The present status of the
excluded regions at short ranges ($z<100\,\mu$m) on the
$(\alpha_{\footnotesize\text{Y}},\lambda_{\footnotesize\text{Y}})$
plane is depicted in figure \ref{FigConstr}.

In the experimental configuration of FORCA-G we have verified that
the only relevant Yukawa-type contribution is the one associated
to the atom-mirror gravitational interaction. At the same time,
the Newtonian part of the atom-surface interaction is completely
negligible with respect to the Earth-atom term already taken into
account in the Wannier-Stark Hamiltonian \eqref{Htot} and with
respect to the expected experimental uncertainties. As a
consequence, the correction we are looking for is obtained by
integrating the Yukawa part of eq. \eqref{YCorr} over the volume
occupied by the surface. Describing the mirror as a cylinder (the
atom being on the direction of its axis) and recalling the we are
looking for deviations having length scale
$\lambda_{\footnotesize\text{Y}}$ in the $\mu$m range we obtain,
after a straightforward calculation,
\begin{equation}\label{HY}H_{\footnotesize\text{Y}}(z)=2\pi\alpha_{\footnotesize\text{Y}}G\rho_{\footnotesize\text{S}}m\lambda_{\footnotesize\text{Y}}^2e^{-\frac{2z}{\lambda_{\footnotesize\text{Y}}}}\end{equation}
$\rho_{\footnotesize\text{S}}$ being the density of the surface.
We are now going to find the new unperturbed energy levels of the
system (in absence of Casimir-Polder interaction) in presence of
the new hamiltonian term \eqref{HY}. This can be done using the
method described in section \ref{Sec:3}, after having chosen the
value of the parameters $\alpha_{\footnotesize\text{Y}}$,
$\lambda_{\footnotesize\text{Y}}$ and
$\rho_{\footnotesize\text{S}}$. The new eigenvalues of the
unperturbed Hamiltonian
$H_{\footnotesize\text{WS}}+H_{\footnotesize\text{Y}}$ will be
noted with $E^{\footnotesize\text{(Y)}}_n$ for each well $n$. As
far as the surface density is concerned, since we still do not
have any information about the surface to be used in the
experiment, we choose throughout this section just as an example
the density of silicon
$\rho_{\footnotesize\text{S}}=2.33\times10^3\,\text{kg}\,\text{m}^{-3}$,
close to the values corresponding to SiO$_2$ or BK7 typically used
in experiments.

As anticipated in the introduction, one of the the scopes of the
experiment FORCA-G is to look for Yukawa-type deviations both near
the surface (at distances of the order of $\mu$m) and in the
region where the Casimir-Polder interaction can be theoretically
modelled at a degree of precision comparable to the experimental
noise. In the former regime, the idea of the experiment is to
compare the results obtained using two different isotopes of
Rubidum (in particular, $^{85}$Rb and $^{87}$Rb) in order to make
the energy differences between wells almost independent on the
Casimir-Polder interaction \cite{WolfPRA07}. As a consequence,
when discussing the Yukawa correction near the surface, we first
need to calculate (both for Wannier-Stark and Yukawa potentials)
the differences in energy levels $E_n$ and
$E^{\footnotesize\text{(Y)}}_n$ between $^{85}$Rb and $^{87}$Rb,
calculated using the formalism described in the previous sections
with the different isotope masses in the Hamiltonians \eqref{Htot}
and \eqref{HY}. These differences will be noted with.
\begin{equation}\mathcal{D}E_n=\Bigl(E^{85}_n-E^{87}_n\Bigr)-\Bigl(E^{\footnotesize\text{(Y)}85}_n-E^{\footnotesize\text{(Y)}87}_n\Bigr)\end{equation}
Finally the experiment will be able to detect a Yukawa-type
deviation if the difference $\mathcal{D}E_n$ is within the
experimental sensitivity. In the case of FORCA-G, the expected
sensitivity is $10^{-4}\,$Hz \cite{WolfPRA07}.

We first give in table \ref{TableEY1} the results obtained for
$\alpha_{\footnotesize\text{Y}}=3\times10^{10}$ and
$\lambda_{\footnotesize\text{Y}}=1\,\mu$m: the value of
$\alpha_{\footnotesize\text{Y}}$ approximately corresponds to the
limit of the experimentally accessed region for
$\lambda_{\footnotesize\text{Y}}=1\,\mu$m.
\begin{center}\begin{table}[h]\begin{tabular}{|c|r|c|c|r|}
\hline
$n$ & $\mathcal{D}E_n$ (Hz) & & $n$ & $\mathcal{D}E_n$ (Hz)\\
\hline\hline
1  & 1.654[-1] & & 13 & 2.0[-3]\\
2  & 1.425[-1] & & 14 & 1.5[-3]\\
3  & 1.221[-1] & & 15 & 1.2[-3]\\
4  & 9.43[-2] & & 16 & 9[-4]\\
5  & 5.72[-2] & & 17 & 7[-4]\\
6  & 2.71[-2] & & 18 & 5[-4]\\
7  & 1.30[-2] & & 19 & 4[-4]\\
8  & 8.0[-3] & & 20 & 3[-4]\\
9  & 6.0[-3] & & 21 & 2[-4]\\
10 & 4.4[-3] & & 22 & 2[-4]\\
11 & 3.4[-3] & & 23 & 1[-4]\\
12 & 2.6[-3] & & 24 & 1[-4]\\
\hline\end{tabular}\caption{First 24 values (the energy
differences are expressed in Hz) of the modified Wannier-Stark
spectrum for $U=3$ and in presence of the Yukawa-type potential
\eqref{HY} with $\alpha_{\footnotesize\text{Y}}=3\times10^{10}$
and
$\lambda_{\footnotesize\text{Y}}=1\mu\text{m}$.}\label{TableEY1}\end{table}\end{center}
From these results it is clear that the Yukawa-type deviations
corresponding to the couple
$(\alpha_{\footnotesize\text{Y}},\lambda_{\footnotesize\text{Y}})$
chosen are, in principle, experimentally detectable up to the well
$n=24$ in a differential $^{85}$Rb\,--\,$^{87}$Rb measurement.

We now turn to the second experimental configuration in which the
Casimir-Polder potential is expected to be predicted at the
$10^{-4}\,$Hz level \cite{WolfPRA07}. We stress that, apart from
the precision in the calculation of the non-regularized
Casimir-Polder potential, we must pay attention to the uncertainty
introduced by our effective description of the finite size of the
atom. Assuming that the Casimir-Polder potential can be
theoretically determined, independently of its regularization,
with at best a $1\%$ accuracy, the absolute precision in its
determination can be considered comparable to the experimental
error already around the well $n=40$, where $z=10\,\mu$m and the
potential equals approximately $0.06\,$Hz, i.e. in this second
experimental configuration the atoms will be at 10\,$\mu$m or more
from the surface. At this distance, our hypothesis of a spherical
atom plays already no role. Indeed, we have checked that, using
both probability distributions and both radii, the potentials so
obtained differ less than $10^{-5}\,$Hz already at $z=5\,\mu$m.
This is coherent with the fact that the finite size of the atom
plays a negligible role at distances much larger than the atomic
size itself. As a consequence, in this second experimental regime
the precision on the standard calculation and the experimental
uncertainties impose stronger limitations than our effective
model.

We have calculated the Yukawa corrections on the well $n=40$ for
different values of $\lambda_{\footnotesize{Y}}$: for each of
them, we have found our limiting value of
$\alpha_{\footnotesize{Y}}$ by looking for a correction of the
order of $10^{-4}\,$Hz. We have, moreover, repeated the same
calculation for $n=70$ (where $V_\text{CP}\simeq0.01\,$Hz) as well
as in the near regime discussed above evaluating the energy
difference between wells $n=4$ and $n=6$. These three curves are
represented in figure \ref{FigConstr} on top of the present
experimental constraints.

\section{Discussion}\label{Sec:6}

In this work, we have introduced an effective model describing the
atom as a spherical probability distribution. This was needed in
order to regularize the expression of the Casimir-Polder energy
correction to the modified Wannier-Stark states. It is important
to discuss in more detail the validity of this model in connection
with the experimental results.

Let us start by recalling that in the context of the search for
non-Newtonian gravitation, our model does not impose severe
limitations. As a matter of fact, in the near regime the
Casimir-Polder contribution is almost cancelled by the use of two
isotopes, whereas at far distances we have shown (see sec.
\ref{Sec:5}) that the error introduced by our description is
negligible with respect to the accuracy in the knowledge of the
Casimir-Polder potential itself.

The experiment could be used, in addition, to test the validity of
our model. To this aim, measurements should be performed in the
near regime (say within the first ten wells) with a single
isotope. In this case, the measured energy differences would check
the consistency of our atomic description as well as provide an
estimation of the effective radius. The use of a single isotope
makes the correction coming from the Yukawa potential negligible
with respect to the Casimir-Polder term (see tables
\ref{TableCorr1} and \ref{TableEY1}).

Finally, the experimental setup can be used for a measurement of
the Casimir-Polder potential around $5\,\mu$m: in this region, as
shown in section \ref{Sec:5}, the energy correction due to the
atom-field interaction is almost insensitive to the model chosen
and the Yukawa interaction is much smaller than the quantum
electrodynamical one. This measure could provide a new
experimental observation of the Casimir-Polder potential with a
relative uncertainty of less than one part in $10^3$.

Nevertheless, we stress here that a precise knowledge of the
Casimir-Polder standard potential requires an accurate description
of the atomic and surface optical data. The details of the latter
are unavailable at present, so the calculation in this paper have
performed for perfectly conducting mirror. To complete our
analysis it will be enlightening to compare our results to the
exact sphere-plate calculations
\cite{CanaguierPRL09,CanaguierArxiv}. In this case, as remarked in
\cite{CanaguierArxiv}, an appropriate description of the
dielectric properties of the sphere is needed to mimic the atomic
optical response.

\section{Conclusions}

In this paper we have discussed the modifications of the
Wannier-Stark states in presence of a surface. As a first step we
have considered the presence of the surface as a boundary
condition of the time-independent Schr\"{o}dinger equation
obtaining in this way a new class of states. These states, even if
asymptotically coincident with the ordinary Wannier-Stark states
at large distances from the surface, significantly differ from
them, both in energy and shape of the wavefunction, at the first
few wells.

We have then also taken into account the Casimir-Polder
interaction between the atom and surface as a source of correction
to the energy levels of the system. We have shown that these
corrections diverge due to the $z^{-3}$ behavior of the
electrodynamical potential energy. In order to regularize this
result, we have introduced an effective description of the atom as
a probability density distributed over a spherical volume. Our
description leaves as free parameters both the radius of the
sphere and the probability distribution. We have characterized the
dependence of our results on both quantities. The validity of this
model as well as the values of these parameters remain to be
investigated by experiments.

In the second part of the paper we have studied the possibility of
measuring a hypothetical Yukawa-type contribution to the
gravitational potential at short distances. We have calculated the
constraints that the experiment FORCA-G will be able to set on the
$(\alpha_{\footnotesize\text{Y}},\lambda_{\footnotesize\text{Y}})$
plane. We have shown that the constraints set by the experiment
are dominated by the experimental uncertainties and unaffected (to
within those uncertainties) by the choice of the model for the
regularization of the CP interaction.

This work paves the way to the precise calculation of the energy
levels in the experimental configuration of FORCA-G and other
experiments that use atoms in optical dipole traps close to a
surface \cite{WolfPRA07,SorrentinoPRA09}. To this aim, a precise
knowledge of the optical data of the mirror and the atom is
needed. This information will allow us to give a more detailed
estimate of the accuracy of our results, also based on the
comparison with independent approaches to the regularization
problem. Finally, the knowledge of the atomic wavefunctions
constitutes the first ingredient for the description of the
dynamics of the system, which is the subject of ongoing work.

\begin{acknowledgments}
This research is carried on within the project iSense, which
acknowledges the financial support of the Future and Emerging
Technologies (FET) programme within the Seventh Framework
Programme for Research of the European Commission, under FET-Open
grant number: 250072. We also gratefully acknowledge support by
Ville de Paris (Emergence(s) program) and IFRAF. The authors thank
Q. Beaufils, A. Canaguier-Durand, R. Guérout, P. Lemonde, R.
Passante, F. Pereira dos Santos and S. Reynaud for fruitful and
stimulating discussions.
\end{acknowledgments}


\begin{thebibliography}{48}
\bibitem{FootnoteCasimir}The Casimir-Polder potential is naturally expressed, as a result of a perturbative calculation,
as an integral over the real axis of frequencies (see e.g. eq.
\eqref{VTZ}). In order to make calculations easier, a rotation to
the imaginary axis is usually performed
\cite{ScheelActaPhysSlov08}. In the case of zero temperature,
since the integrand function has no poles on both axes, the result
is an integral over imaginary frequencies \eqref{Vz0}. On the
contrary, the integrand function in the case of non-zero
temperature has an infinite number of poles on the imaginary axis,
namely the Mastubara frequencies $\xi_n=2\pi nk_BT/\hbar$. This
explains the sum over $n$ appearing in eq. \eqref{Vz}.
\bibitem{FootnotePolarizability}It may be useful to specify that
the polarizability $\alpha(\omega)$ defined in eq. \eqref{Polar}
is given in SI units and has the dimensions of C\,m$^2$\,V$^{-1}$.
The corresponding expression in cgs units is given by
$\frac{\alpha(\omega)}{4\pi\epsilon_0}$ and has the dimensions of
a volume.
\end{thebibliography}
\end{document}